\documentclass[usenatbib]{emulateapj}
\usepackage{graphicx}
\usepackage[flushleft]{threeparttable}
\usepackage[usenames,dvipsnames,svgnames,table]{xcolor}
\usepackage{hyperref}
\definecolor{darkblue}{rgb}{0.0,0.0,0.3}
\hypersetup{colorlinks,breaklinks,
            linkcolor=darkblue,urlcolor=darkblue,
            anchorcolor=darkblue,citecolor=darkblue}
\usepackage{amsmath,amssymb}
\usepackage{amsmath}

\usepackage{natbib}
\makeatletter
\renewcommand{\p@subsection}{}
\renewcommand{\p@subsubsection}{}
\makeatother

\begin{document}

\def\etal{et al.\ \rm}
\def\ba{\begin{eqnarray}}
\def\ea{\end{eqnarray}}
\def\etal{et al.\ \rm}
\def\Fdw{F_{\rm dw}}
\def\Tex{T_{\rm ex}}
\def\Fdis{F_{\rm dw,dis}}
\def\Fnu{F_\nu}
\def\WD{\rm WD}

\newcommand\cmtrr[1]{{\color{red}[RR: #1]}}


\title{Non-Gravitational Forces and Spin Evolution of Comets}

\author{Roman R. Rafikov\altaffilmark{1,2}}
\altaffiltext{1}{Centre for Mathematical Sciences, Department of Applied Mathematics and Theoretical Physics, University of Cambridge, Wilberforce Road, Cambridge CB3 0WA, UK; rrr@damtp.cam.ac.uk}
\altaffiltext{2}{Institute for Advanced Study, Einstein Drive, Princeton, NJ 08540}


\begin{abstract}
Motion of many comets is affected by non-gravitational forces caused by outgassing from their surfaces. Outgassing also produces reactive torques resulting in cometary spin evolution. We propose that the two processes are correlated and show that the change of cometary spin rate over its heliocentric orbit scales linearly with the amplitude of its non-gravitational acceleration. The proportionality constant depends on the comet size and orbital elements (semi-major axis and eccentricity) and on the (dimensionless) lever arm parameter $\zeta$ that relates the outgassing-induced torque and acceleration. We determine $\zeta$ for 7 comets for which both non-gravitational acceleration and change of spin period $\Delta P$ were measured and verify this relation. This sample spanning almost 4 decades in $\Delta P$ yields $\log\zeta=-2.21\pm 0.54$,  surprisingly small value and spread. We then apply our framework to 209 comets with measured non-gravitational accelerations and determine the objects most likely to exhibit large spin period changes, $\Delta P\gtrsim 20$ min per orbit assuming rotation period of 10 hr and $\zeta$ comparable to our control sample. These objects should be primary targets for future studies of cometary spin variability, further constraining  distribution of $\zeta$. Lack of comets with very high expected spin rate changes (which is not equivalent to having the highest non-gravitational acceleration) suggests that (1) cometary fission due to outgassing-driven spin-up must be an important process and (2) the distribution of $\zeta$ has a lower limit $\sim 10^{-3}$.
\end{abstract}




\section{Introduction.}  
\label{sect:intro}


Motion of many Solar System bodies is known to be affected by non-gravitational forces, which are not accounted for by the combined effect of the Solar gravity and planetary perturbations \citep{Whipple1950,Sekanina1981,Yeomans2004}. These forces exhibit themselves through changes of orbital elements of minor objects, and are naturally explained as the reactive (jet) forces due to outgassing --- loss of volatiles from the surface --- powered by Solar heating \citep{Bessel,Whipple1950,Marsden1968,Marsden1969}. Non-gravitational acceleration $\boldsymbol{a}_{\rm ng}$ has been measured for several hundred Solar System objects, both comets \citep{Krolik,Szut} and asteroids \citep{Hui2017}. 

Loss of volatiles should also affect the spin angular momentum of the outgassing body. Indeed, there is no reason for the mass loss from the surface of any minor object to be perfectly symmetric. Given the complicated shapes of cometary nuclei, it is natural to expect the reactive forces caused by ejection of volatile material from the surface to not point through its center of mass in general. As a result, reactive forces must affect not only the translational motion of an object but also its spin via the associated torques \citep{Sekanina1981,Jewitt1997,Samar2007}. Net torque acting on a body can change the direction of its spin axis (resulting in forced precession) as well as its spin rate. 

Common underlying origin --- outgassing --- suggests that non-gravitational acceleration and spin evolution of comets should be correlated in some way. This possibility was explored by \citet{Whipple} who showed (using a particular outgassing model) that the rate of forced precession of the cometary spin axis should be proportional to the amplitude of the non-gravitational acceleration $a_{\rm ng}$. In turn, spin precession may cause secular variation of the net non-gravitational linear acceleration, since in many models $\boldsymbol{a}_{\rm ng}$ depends on the geometry of Solar illumination (i.e. incidence angle with respect to local normal) of the cometary surface \citep{Whipple,Sekanina1984}.  

Of course, in practice it is very difficult to determine observationally the direction of cometary spin axis \citep{Sekanina1981,Jewitt1997,Bair}, not even mentioning its variation in time caused by forced precession. On the other hand, it is much easier to measure the {\it rotation rate} of a minor object and its variation. Indeed, changes of spin period have been measured for several Solar System comets \citep{Mueller1996,Knight,Mottola,Bode}. And since period variation is just another manifestation of spin angular momentum evolution (complementary to forced precession\footnote{\citet{Whipple} modeled cometary nucleus as an oblate spheroid with outgassing force pointing through the rotation axis at every point on the surface. This geometry does not result in changes of the spin rate. The fact that cometary spin rate variations are observed (\S \ref{sect:zeta}) implies that this model is too simplistic.}), one expects it to also be related to the net non-gravitational acceleration in some way.

In this work we use simple heuristic arguments to predict a direct correlation between the changes of the cometary spin period and  non-gravitational acceleration acting on them (\S \ref{sect:spin}). We then use observations of the spin rate variability for a sample of comets to confirm this correlation and to determine its characteristics, namely, the value of the lever arm relating the non-gravitational acceleration to the torque it produces (\S \ref{sect:zeta}). We then use this correlation to assess spin period variability for a larger sample of comets with measured non-gravitational accelerations (\S \ref{sect:spin_comets}). We discuss the implications of our results in \S \ref{sect:disc}.


\section{Connecting non-gravitational acceleration and spin evolution}  
\label{sect:spin}


We consider a very general picture of outgassing, in which a specific reactive force per unit area ${\bf f}$ acts on every surface element $dS$ of the body. In general, ${\bf f}$ does not have to be normal to the local surface and it could be zero away from the active sites of mass loss. The full non-gravitational force acting on an object is
\ba  
{\bf F}_{\rm ng}=\int_S {\bf f}~dS,
\label{eq:force}
\ea   
where integration is carried over the full surface. In general, ${\bf F}_{\rm ng}$ varies in time in the inertial frame because of object's rotation. Only the component along the instantaneous spin direction ${\bf \Omega}$ does not average out to zero and gives rise to the observed anomalous acceleration.

At the same time, the net torque acting on an object is
\ba
{\bf T}=\int_S {\bf R}\times {\bf f}~dS,
\label{eq:torque}
\ea
where ${\bf R}$ is the radius-vector from the body center of mass to a surface element $dS$. Note that in the body frame the torque ${\bf T}$ is invariant with regard to object's rotation. Also, while the intensity of outgassing (and value of $|{\bf f}|$) definitely vary along the heliocentric orbit, the direction of elementary torque ${\bf R}\times {\bf f}$ likely does not change much. 

Non-zero torque leads to evolution of the spin angular momentum of the body ${\bf S}$ through $\dot {\bf S}={\bf T}$, resulting in the change of both the direction of ${\bf S}$ (forced precession) as well as its magnitude. The former was explored by \citet{Whipple}, and we focus on the latter in this work. For our present purposes, the evolution of the spin rate $\Omega=|{\bf \Omega}|$ can be adequately described as \citep{Samar}
\ba  
I\dot\Omega=T_\Omega,
\label{eq:spin_ev}
\ea
where $T_\Omega={\bf T}\cdot{\bf \Omega}/\Omega$ is the magnitude of torque ${\bf T}$ along ${\bf \Omega}$ and $I$ is the object's moment of inertia with respect to the ${\bf \Omega}$ axis (the other components of ${\bf T}$ lead to forced precession of the object's spin).

In general, the integrals in equations (\ref{eq:force}) and (\ref{eq:torque}) cannot be directly expressed through one another. However, given their form, it is natural to expect the torque $T=|{\bf T}|$ acting on the object to be related to the magnitude of its non-gravitational force $F_{\rm ng}=|{\bf F}_{\rm ng}|$ via some "effective lever arm". In other words, we should be able to write
\ba   
T_\Omega=\zeta D F_{\rm ng},
\label{eq:F}
\ea
where $\zeta D$ is the effective lever arm (with the dimension of length), $D$ is the characteristic size of the object (e.g. radius for a roughly spherical object), and  $\zeta$ is the dimensionless "lever arm" parameter. 

Coefficient $\zeta$ accounts for the non-central nature of mass loss from the object's surface and provides a simple connection between $T_\Omega$ and $F_{\rm ng}$. A purely central mass loss (with ${\bf f}$ passing through the center of mass at any point on the surface) would result in $\zeta=0$ --- i.e. no torque, but a non-zero force in general. At the other extreme, net force averaging to zero may still give finite $T_\Omega$, meaning $\zeta\to \infty$. However, for a random distribution of multiple outgassing sites over a non-spherical surface one should probably expect $\zeta\lesssim 1$.

Obviously, given $T_\Omega$ and $F_{\rm ng}$ one can always define $\zeta$ simply as a dimensionless coefficient relating the two via equation (\ref{eq:F}). However, it is also natural to expect that in a large sample of objects $\zeta$ may be distributed in a certain way, which eventually reflects (possibly in a rather complicated fashion) the geometry of outgassing, distribution of object shapes, number of active sites, etc. In the following (\S \ref{sect:zeta}) we will operate based on this premise and will use observations to determine the gross characteristics of such a distribution of $\zeta$.

Equations (\ref{eq:spin_ev}) and (\ref{eq:F}) can be combined into
\ba  
\dot\Omega=\zeta\frac{D M}{I}a_{\rm ng},
\label{eq:spinev}
\ea
where $a_{\rm ng}=|\boldsymbol{a}_{\rm ng}|$ is the magnitude of the non-gravitational acceleration defined through ${\bf F}_{\rm ng}=M\boldsymbol{a}_{\rm ng}$. This relation illustrates the expectation of a direct correlation between the variation of the spin rate (or period) of a minor object and the magnitude of its outgassing-induced linear non-gravitational acceleration $a_{\rm ng}$. Note that, unlike the forced precession rate\footnote{Which obeys a relation very similar to (\ref{eq:spinev}).} derived by \citet{Whipple}, $\dot\Omega$ is independent of the spin period of an object. 

\begin{table*}
\begin{threeparttable}
\caption{Solar System objects with measured non-gravitational acceleration and spin period changes}
\begin{tabular}{lcccccccccc}
    \hline \hline\\
    Name & $a$ & $e$ & $R$ & $(A_r,A_\varphi,A_n)$ & $P$ & $\Delta P$/orbit & $\Delta \Omega$/orbit & $\Delta \Omega_1$ & $\zeta$ & Ref.
    \\ 
     & $\big[$au$\big]$ &  & $\big[$km$\big]$ & $\big[10^{-10}$ au d$^{-2}\big]$ & $\big[$hr$\big]$ & $\big[$min$\big]$ & $\big[$s$^{-1}\big]$ & $\big[$s$^{-1}\big]$ &  & \\
    \\
    \hline
    \\
2P/Encke &	2.2	& 0.85 & 2.4 & $\left(-0.4^{\pm 0.47},-0.07^{\pm 0.03},0.0\right)$\footnotemark[1] & 11 & 4 & $9.6\times 10^{-7}$ & $3.04\times 10^{-4}$  & 0.031 & 1,2 \\ 
9P/Tempel 1 & 3.15 & 0.51 & 2.8	& $\left(-31^{\pm 14},-0.2^{\pm 8.8}, -4.2^{\pm 2.1}\right)$ & 41 & -14 & $2.4\times 10^{-7}$ & $3.62\times 10^{-4}$ & $6.7\times 10^{-4}$ & 1,2,3\\
10P/Tempel 2 & 3.07	& 0.54 & 6.0 & $\left(2.3^{\pm 0.3},0.1^{\pm 0.02}, 1.8^{\pm 0.3}\right)$ & 9 & 0.27\footnotemark[2]  & $9.7\times 10^{-9}$ & $1.93\times 10^{-5}$ & 0.005 & 1,2,4,5\\
19P/Borrelly & 3.6 & 0.62 & 2.5	& $\left(19^{\pm 0.2},-0.78^{\pm 0.07},2.8^{\pm 0.3}\right)$ & 28 & 40 & $1.5\times 10^{-6}$ & $3.0\times 10^{-4}$  & 0.0048 & 2\\
67P/C-G\footnotemark[3] & 3.46	& 0.64 & 1.65 &	$\left(11^{\pm 0.2},-0.37^{\pm 0.05},2.5^{\pm 0.16}\right)$ & 12 & -21 & $4.4\times 10^{-6}$ & $3.2\times 10^{-4}$  & 0.014 & 2,6\\
103P/Hartley 2 & 3.47 & 0.7	& 0.58 & $\left(7.6^{\pm 0.5},2.5^{\pm 0.02},0.0\right)$ & 18 & 150 & $1.2\times 10^{-5}$ & $8.3\times 10^{-4}$  & 0.014 & 1,2\\
41P/T-G-K\footnotemark[4] & 3.08 & 0.66 & 0.7 & $\left(170^{\pm 2},43^{\pm 2},14^{\pm 1}\right)$ & 20 & $>1560$ & $4.9\times 10^{-5}$ & $1.56\times 10^{-2}$  & 0.003 & 7 \\ 
\\
    \hline   
\end{tabular}
\begin{tablenotes}
\item {\bf Notes}: For every comet we list semi-major axis and eccentricity of its heliocentric orbit, effective physical radius, 3 components of the non-gravitational acceleration at 1 au (assuming \citet{Marsden} model (\ref{eq:gr})), rotation period, its change over an orbit, change of spin rate over an orbit, $\Delta\Omega_1$ defined by equation (\ref{eq:dOmmax}), and lever arm parameter $\zeta$ defined by equation (\ref{eq:F}).\\  
Key to references: (1) \citet{Samar}, (2) \citet{Mueller}, (3) \citet{Belton2011}, (4)  \citet{Mueller1996}, (5) \citet{Knight}, (6) \citet{Mottola}, (7) \citet{Bode}. Data on $(A_r,A_\varphi,A_n)$ come from JPL Small Body Database (https://ssd.jpl.nasa.gov/?comets)
\footnotetext[1]{For 2P/Encke Small Body Database does not compute $A_n$, so we set it to zero. Also, a non-standard form of $g(r)$ is used in the database. For that reason we used $A_\varphi=-7\times 10^{-12}$ au d$^{-2}$ from \citet{Sosa}, which is based on the conventional \citet{Marsden} model, and scaled up the value of $A_r$ from the database correspondingly.}
\footnotetext[2]{For this object Table 3 of \citet{Bode} cites incorrect value of $\Delta P/$orbit: 2.7 min instead of the correct 0.27 min \citep{Mueller1996,Knight}.}
\footnotetext[3]{Comet 67P/Churyumov-Gerasimenko.}
\footnotetext[4]{Comet 41P/Tuttle-Giacobini-Kres\'ak.}
\end{tablenotes}
\label{table:zeta}
\end{threeparttable}
\end{table*}


\section{Determination of the lever arm parameter $\zeta$ based on the Solar System data}  
\label{sect:zeta}


In this section we verify the correlation (\ref{eq:spinev}) between $a_{\rm ng}$ and spin evolution using information about spin period changes of some Solar System objects. This exercise also allows us to estimate the typical value of the dimensionless parameter $\zeta$. 

There are more than 200 comets, for which the non-gravitational acceleration measurements are reported in the JPL Small Body Database\footnote{https://ssd.jpl.nasa.gov/?comets}. However, the changes of the spin period $\Delta P$ (or spin rate $\Delta\Omega$) have been measured for only several comets. Seven of these objects, listed in Table \ref{table:zeta}, also have measurements of their non-gravitational acceleration available to us, making possible the determination of $\zeta$. There are several other comets for which rotation period variations were reported, e.g.  C/2001 K5 (LINEAR) \citep{Drahus2006} and C/1990 K1 (Levy)\footnote{Also known as Comet Levy (1990c).} \citep{Schleicher,Feldman}, however, we were not able to find the non-gravitational acceleration data for them. 

For simplicity we treat the comets listed in Table \ref{table:zeta} as spheres of uniform density with effective radius $D=R$, so that their moment of inertia is $I=(2/5)MR^2$. Then one can use equation (\ref{eq:spinev}) to write the change of $\Omega$ accumulated over the full heliocentric orbit of an object as
\ba  
\Delta\Omega=5\zeta R^{-1}\int_{r_p}^{r_a}a_{\rm ng}(r)\frac{dr}{v_r},
\label{eq:dOm}
\ea  
where $r_p=a(1-e)$ and $r_a=a(1+e)$ are the perihelion and aphelion distances, correspondingly, and $v_r$ is the heliocentric radial velocity. 

Clearly, $\Delta\Omega$ depends on the radial profile of the non-gravitational acceleration $a_{\rm ng}(r)$. 
It is customary to parametrize $\boldsymbol{a}_{\rm ng}(r)$ using a simple model 
\ba  
\boldsymbol{a}_{\rm ng}(r)={\bf A}g(r),~~~~~{\bf A}=(A_r,A_\varphi,A_n),
\label{eq:a_ng}
\ea  
where the dimensionless function $g(r)$ describes the radial dependence of the non-gravitational acceleration. It is customary to define $g(r)$ such that $g(1~$au$)=1$. Constant vector ${\bf A}$ sets the amplitude of acceleration vector; it has components $A_r$, $A_\varphi$ in the radial and azimuthal directions, correspondingly, in the plane of the orbit, and $A_n$ normal to it; we depart from the conventional notation ${\bf A}=(A_1,A_2,A_3)$.  

Many different prescriptions have been proposed for the scaling function $g(r)$ \citep{Yeomans2004}. The most commonly used expression for $g(r)$, motivated by the expected water sublimation rate powered by Solar heating, is due to \citet{Marsden}: 
\ba  
g(r)=\alpha\left(\frac{r}{r_0}\right)^{-m}\left[1+\left(\frac{r}{r_0}\right)^{n}\right]^{-k},
\label{eq:gr}
\ea 
where $\alpha=0.1113$, $m=2.15$, $n=5.093$, $k=4.6142$, and $r_0=2.808$ au, so that $g(1~$au$)=1$. Note that this expression assumes water vapor production to be symmetric with respect to the perihelion of the orbit, which may not be accurate because of thermal inertia. There is a number of refinements to the prescription (\ref{eq:gr}), e.g. accounting for the time lag in water vapor production \citep{Yeomans1989}, secular drift of ${\bf A}$ \citep{Aksnes}, changing orientation of the cometary spin axis \citep{Whipple,Sekanina1981}, sublimation of other species, i.e. not just water \citep{Sekanina2015}. 

Despite these improvements, in this work we employ the conventional prescription (\ref{eq:gr}) for computing $\Delta\Omega$. The main reason being that the non-gravitational measurements reported in the Small Body Database assume this model, and we want to be internally consistent in this regard.


\begin{figure}
\centering
\includegraphics[width=0.48\textwidth]{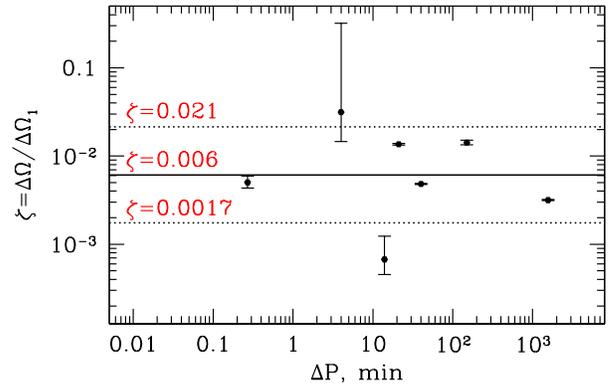}
\caption{
Values of the dimensionless lever arm parameter $\zeta$ defined by equation (\ref{eq:F}) for 7 Solar System comets with simultaneously measured period change (over an orbit) $\Delta P$ and the amplitude of non-gravitational acceleration. Calculation is based on equations (\ref{eq:dOm1})-(\ref{eq:dOmmax}). Horizontal solid and dashed lines correspond to the mean of $\log\zeta$ and $1\sigma$ standard deviation from it (corresponding values of $\zeta$ are labeled in red). 
\label{fig:zeta}}
\end{figure}


For an elliptical Keplerian orbit with a semi-major axis $a$ one has
\ba  
v_r=\left(\frac{GM_\odot}{a}\right)^{1/2}r^{-1}\sqrt{(r_a-r)(r-r_p)}.
\label{eq:vr}
\ea 
Plugging this expression into equation (\ref{eq:dOm})
we find
\ba  
\Delta\Omega = \zeta\Delta\Omega_1,
\label{eq:dOm1}
\ea  
where
\ba  
\Delta\Omega_1 =
5\frac{A}{R}\sqrt{\frac{a}{GM_\odot}}\int_{r_p}^{r_a}\frac{rg(r)dr}{\sqrt{(r_a-r)(r-r_p)}}
\label{eq:dOmmax}
\ea  
is the characteristic value of  $\Delta\Omega$ reached when the lever arm parameter $\zeta=1$. Also, we defined $A=\sqrt{A_r^2+A_\varphi^2+A_n^2}$ to account for the fact that all three components of the reactive force ${\bf F}_{\rm ng}$ acting on the comet can lead to its spin evolution. 


\begin{figure*}
\centering
\includegraphics[width=0.99\textwidth]{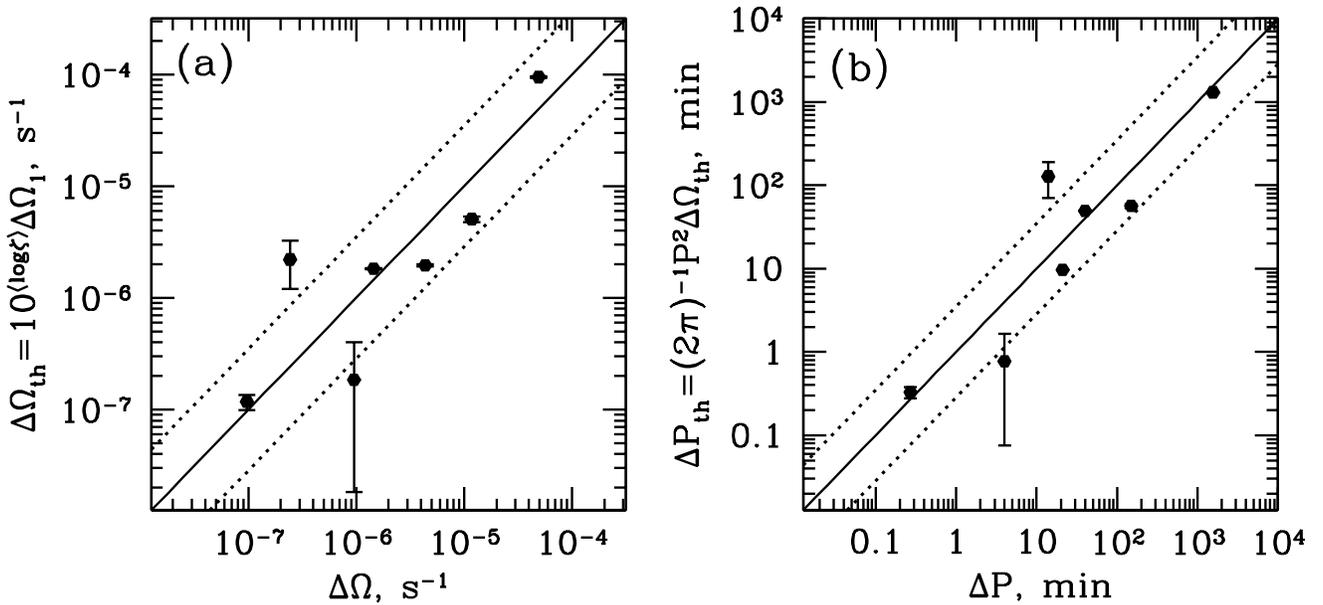}
\caption{
Comparison of the observed changes of (a) spin rate $\Delta\Omega$ and (b) period $\Delta P$ of the comets listed in Table \ref{table:zeta} with the corresponding theoretical predictions based on equations (\ref{eq:dOm1})-(\ref{eq:dOmmax}), assuming lever arm parameter $\zeta=0.006$ (corresponding to $\langle\log\zeta\rangle$ in equation (\ref{eq:stats})). Solid and dotted lines correspond to the mean and standard deviation of $\log\zeta$ in the relation (\ref{eq:dOm1}). Note a strong correlation between the observed and theoretically predicted values over several decades in $\Delta\Omega$ and $\Delta P$.
\label{fig:correlations}}
\end{figure*}


We can now determine the values of the lever arm parameter $\zeta$ for every comet listed in Table \ref{table:zeta}. Using the known orbital elements $a$ and $e$ of these objects, as well as the information on their sizes $R$ and non-gravitational acceleration components ${\bf A}$, we compute $\Delta\Omega_1$ using equation (\ref{eq:dOmmax}). We then use equation (\ref{eq:dOm1}) and observed change of the spin rate $\Delta\Omega$ to find the value of $\zeta=\Delta\Omega/\Delta\Omega_1$ for each comet. 

Results of this exercise are listed in Table \ref{table:zeta} and are also shown in Figure \ref{fig:zeta}. The error bars reflect only the uncertainty in the determination of ${\bf A}$. Uncertain size and shape of many of these comets would introduce additional systematic uncertainty at the level of tens of per cent. At the same time, it is important to note that our determination of $\zeta$ is independent of the cometary bulk density, which is one of the particularly poorly constrained characteristics \citep{Sosa,Sosa2011}.

One can see that, as expected, $\zeta<1$. Also, the values of $\zeta$ show substantial variation (although the most dramatic outliers also have the biggest errors in the determination of their $a_{\rm ng}$). However, one has to keep in mind that they are computed for objects with $\Delta\Omega$ spanning almost 4 orders of magnitude.

Considering these values of $\zeta$ as a realization of an underlying distribution of $\log\zeta$, we find the mean and standard deviation of this distribution to be
\ba  
\langle\log\zeta\rangle=-2.21,~~~\sigma_{\log\zeta}=0.54.
\label{eq:stats}
\ea  
Values of $\zeta$ corresponding to the mean of $\log\zeta$, and deviating from it by $\pm\sigma_{\log\zeta}$ are $\zeta=0.006$, $0.0017$, and $0.021$, correspondingly.

In Figure \ref{fig:correlations} we plot the observed $\Delta\Omega$ against the "mean theoretically expected" value  $\Delta\Omega_{\rm th}=10^{\langle\log\zeta\rangle}\Delta\Omega_1$, computed using equations (\ref{eq:dOmmax}) and (\ref{eq:stats}). We also do the same for the observed change in period $\Delta P$ and the "mean theoretically expected" value\footnote{The small $\Delta\Omega/\Omega$ assumption used in this expression is violated for comet 41P/Tuttle-Giacobini-Kres\'ak, see Table \ref{table:zeta} and \citet{Bode}.} $\Delta P_{\rm th}=P^2\Delta\Omega_{\rm th}/(2\pi)$. Solid and dotted lines in this Figure reflect the mean and standard deviation of $\log\zeta$ in the relation (\ref{eq:dOm1}).

Figure \ref{fig:correlations} clearly illustrates a strong positive correlation between $\Delta\Omega$ and $\Delta\Omega_{\rm th}$ (or $\Delta\Omega_1$); the Spearman's rank correlation coefficient is 0.786. This dataset is consistent with the linear dependence predicted by equations (\ref{eq:dOm1})-(\ref{eq:dOmmax}), thus fully supporting our hypothesis of a direct connection between the non-gravitational acceleration and torques acting on outgassing objects, see equation (\ref{eq:F}). This gives us confidence in extending this idea to other minor objects.


\section{Expected spin evolution of comets}  
\label{sect:spin_comets}



\begin{figure*}
\centering
\includegraphics[width=0.95\textwidth]{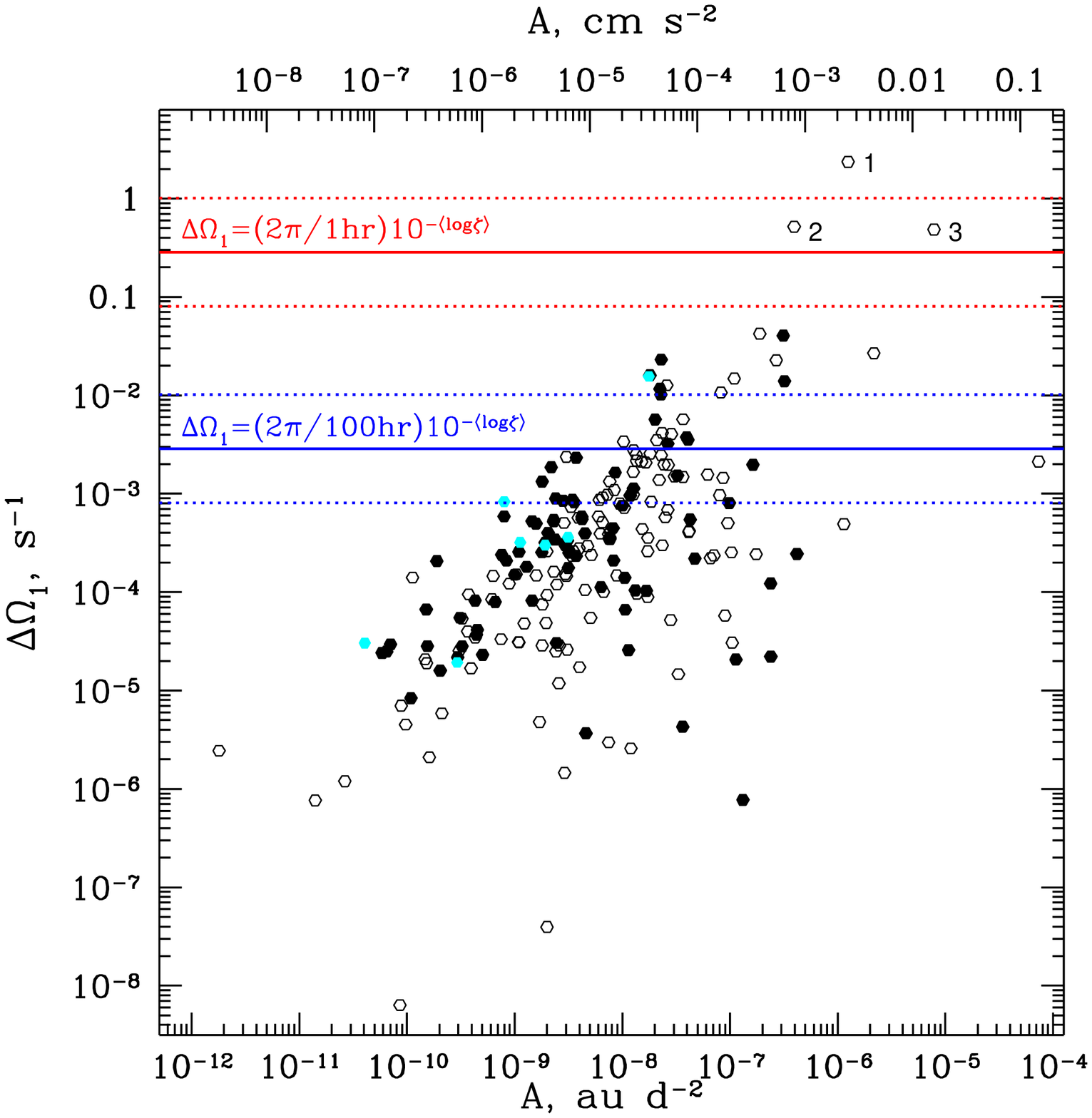}
\caption{
Change of the spin rate $\Delta\Omega_1$ due to non-gravitational forces assuming lever arm parameter $\zeta=1$ accumulated over a single orbit, computed for a sample of 209 Solar System comets with available data on the amplitude of the non-gravitational acceleration (i.e. with measurable ${\bf A}$). Filled hexagons represent objects with measured sizes; empty ones have no size information available and $R=10$ km is set for them. Cyan points represent 7 objects used in \S \ref{sect:zeta} for determining $\zeta$. Horizontal solid curves correspond to rotation rate changes $\Delta\Omega$ of $2\pi/1$ hr (red) and $2\pi/100$ hr (blue) for $\zeta=0.006$ (corresponding to the mean of $\log\zeta$ as determined in \S \ref{sect:zeta}); dashed lines illustrate $1\sigma$ deviation in $\log\zeta$. Three numbered objects are discussed in \S \ref{sect:comets}. See text for more details.
\label{fig:comets}}
\end{figure*}


As mentioned before, determinations of spin period $P$ (or its variation $\Delta P$) are not available for the majority of comets with measured non-gravitational accelerations. This means that we cannot determine the value of the dimensionless lever arm $\zeta$ for these objects. Nevertheless, for each comet we can still use equation (\ref{eq:dOmmax}) to compute its $\Delta\Omega_1$ --- the change of spin frequency per orbit if the lever arm parameter $\zeta$ were equal to unity. Knowing $\Delta\Omega_1$ and assuming that the determinations of $\zeta$ in \S \ref{sect:zeta} reflect the true distribution of the lever arm parameter for all comets, we can then {\it predict} the expected variation of the spin rate per orbit for all comets with measured $\boldsymbol{a}_{\rm ng}$. In some sense, this procedure is inverse to what has been done in \S \ref{sect:zeta}. 
To perform this calculation, we used the Small Body Database to extract a sample of 209 Solar System objects with measured components of ${\bf A}$ in at least one direction. We used their orbital parameters listed in the Database, as well as the information on their sizes, when available, to compute $\Delta\Omega_1$ using equation (\ref{eq:dOmmax}). When no information on size was available, we arbitrarily set $R=10$ km, which likely leads to an underestimate of $\Delta\Omega_1$. Note that this calculation does not use information about the rotational periods of these comets, which are not known in most cases.

Results of this exercise are presented in Figure \ref{fig:comets}, where we plot $\Delta\Omega_1$ vs $A=|{\bf A}|$ for each of these objects (using different symbols for objects with and without size determinations). This Figure reveals a general trend of higher $\Delta\Omega_1$ for objects with larger $A$, which is to be expected since $\Delta\Omega_1\propto A$. However, there is a large scatter around a simple linear relation due to the disparity of orbital and physical parameters of individual objects, which also enter equation (\ref{eq:dOmmax}).

\begin{figure}
\centering
\includegraphics[width=0.5\textwidth]{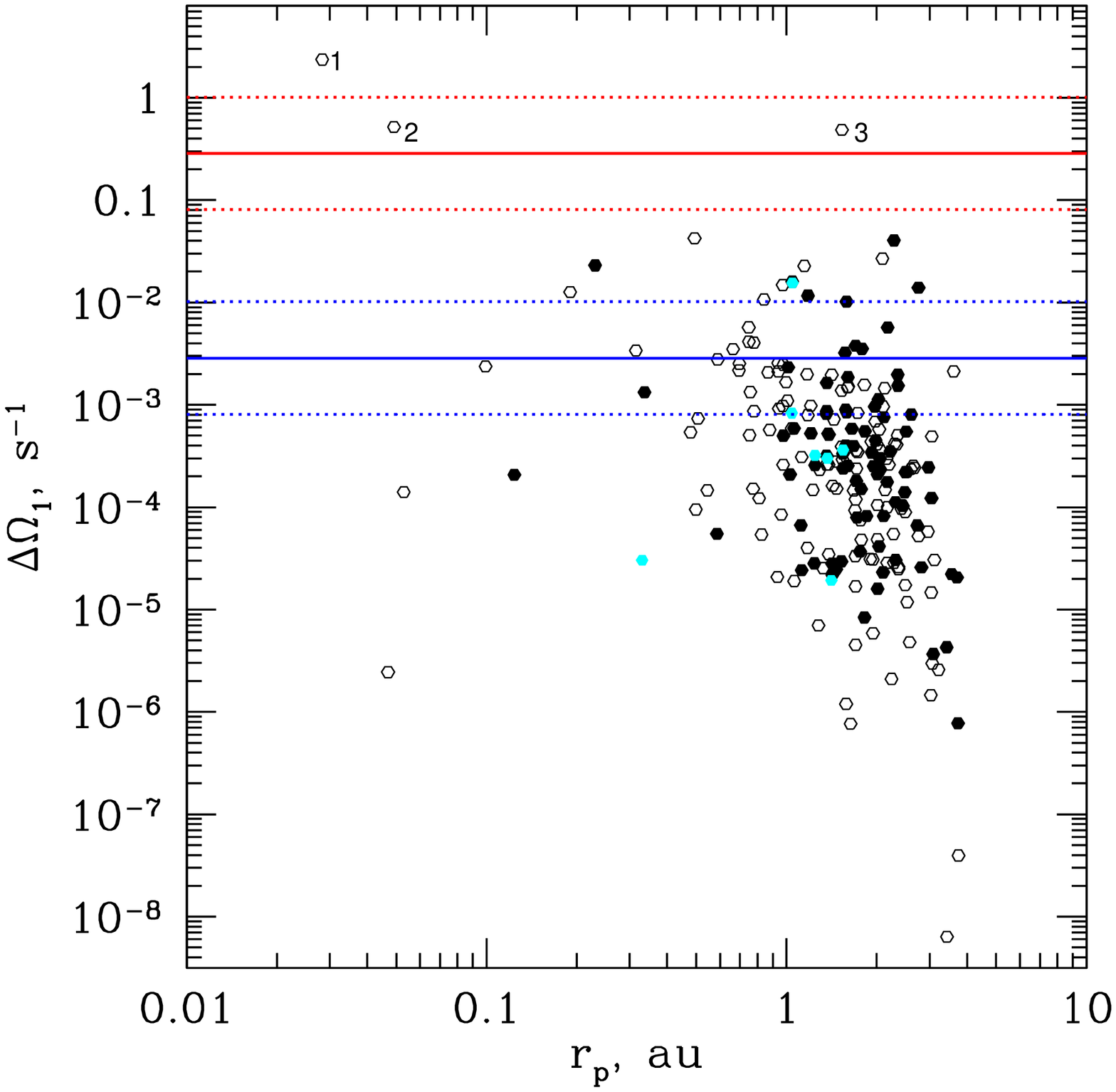}
\caption{
Same as Figure \ref{fig:comets}, but now with $\Delta\Omega_1$ plotted vs perihelion distance $r_p$. Meaning of the horizontal lines, labels, etc. is the same as in Figure \ref{fig:comets}. See text for discussion.
\label{fig:comets_peri}}
\end{figure}

In Figure \ref{fig:comets_peri} we plot $\Delta\Omega_1$ for the same objects but now as a function of their periastron distance $r_p$. There seems to be a general trend of objects with higher $r_p$ to have lower $\Delta\Omega_1$, which is expected because of the steep radial dependence of $g$ on $r$, see equations (\ref{eq:gr}) and (\ref{eq:dOmmax}).


There are two sets of horizontal lines in Figures \ref{fig:comets} and \ref{fig:comets_peri}. The upper (red) solid curve (at $\Delta\Omega_1\approx 0.3$ s$^{-1}$) corresponds to spin-up to period $P_{\rm crit}=1$ hr from a non-rotating state, assuming a lever arm parameter $\zeta=10^{\langle\log\zeta\rangle}\approx 0.006$, characteristic of the comets used in \S \ref{sect:zeta}. The dotted lines mark a standard deviation (in $\log\zeta$) from this value of $\zeta$, see equation (\ref{eq:stats}). We do not expect comets to survive at spin periods $\lesssim P_{\rm crit}$ because {\it rotational fission} would destroy such rapidly spinning objects \citep{Davidsson}. Fragmentation of comets is not an uncommon phenomenon \citep{SekaninaFrag,Ishiguro,Steckloff,Jewitt2016}, although its precise causes are still debated. 

Lower (blue) curves correspond to $10^2$ times lower $\Delta\Omega$ and $\Delta\Omega_1$ (for the same assumptions regarding $\zeta$). This value of $\Delta\Omega$ corresponds to e.g. a period change of $|\Delta P|=|2\pi\zeta P^{-2}\Delta\Omega_1|=1$ hr per orbit for an object initially spinning with $P=10$ hr period (which is rather typical for comets, see Table \ref{table:zeta}). Objects above this set of curves should accumulate measurable changes of their spin periods in a single orbit, even if the uncertainty of period measurement is about an hour. One can see that a substantial number of Solar System comets belong to this category.

In Table 2 we list all comets for which our calculations give $\Delta\Omega_1>10^{-3}$ s$^{-1}$, translating into $|\Delta P|>20$ min per orbit for initial $P=10$ hr and $\zeta=0.006$. In reading this table it should be remembered that many objects do not have reliable size measurements; we set $R=10$ km for them, which likely results in underestimating the true value of $\Delta P$. Also, measurements of ${\bf A}$ often have substantial uncertainty. Nevertheless, these comets are still expected to be good targets for future measurements of their spin variability.  


\subsection{Comets with highest $\Delta\Omega_1$}  
\label{sect:comets}


Out of the full sample of comets used in this calculation, three objects with the highest $\Delta\Omega_1$ end up lying very close (or above) the breakup line in Figure \ref{fig:comets}. Here we look at their properties in some more detail to understand how their existence can be brought in accord with the expectation of their rotational fission. Note that we used $R=10$ km for these objects; if they are smaller, then $\Delta\Omega_1$ would go up making rotational breakup even more of an issue.

{\bf C/2015 D1 (SOHO)}  This comet (marked $1$ in Figure \ref{fig:comets}) has $a=4.94$ au, $e=0.99427$, corresponding to perihelion distance $r_p=0.0283$ au. Because of the low $r_p$ its $\Delta\Omega_1=2.36$ s$^{-1}$ is the highest among the Solar System comets in our sample. This Sun-skirting comet has, in fact, not survived its perihelion passage in 2015 \citep{Hui}. It is plausible that rotationally-induced stresses have contributed to its destruction.

{\bf P/1999 J6 (SOHO)} This Jupiter-family comet ($2$, $a=3.1$ au, $e=0.9842$) is also a Sun-skirting object with perihelion at $r_p=0.0491$ au; it has the second-highest $\Delta\Omega_1=0.52$ s$^{-1}$. Despite its low $r_p$ and high $\Delta\Omega_1$, it has been observed over several orbits, which are clearly affected by the non-gravitational effects (its dominant radial acceleration component $A_1$ is determined with $40\%$ accuracy). Most likely explanations for why it has so far survived rotational fission are (1) either a very low value of $\zeta\lesssim 10^{-3}$ characterizing the non-gravitational torque acting on it, or (2) poor characterization of its non-gravitational force behavior (especially near perihelion, where most of $\Delta\Omega_{\rm max}$ gets accumulated) by the equation (\ref{eq:gr}), see \citet{Sekanina2015}.

{\bf 205P/Giacobini-B} Unlike the previous two objects, this Jupiter-family comet ($3$, $a=3.6$ au, $e=0.575$) has rather large perihelion distance $r_p=1.53$ au. Its high $\Delta\Omega_1=0.49$ s$^{-1}$ is caused by large $A$. However, the value of $A$ listed in the Small Body Database has been determined based on only 19 d long orbital arc. This raises a concern about the accuracy of the non-gravitational acceleration measurement for this particular object.


\section{Discussion}  
\label{sect:disc}


Our finding of a clear correlation between the anomalous (non-gravitational) linear acceleration of some minor objects and variation of their rotational period (\S \ref{sect:zeta}) has important implications for Solar System comets, which we discuss in more detail in \S \ref{sect:disc_SS}.

The existence of this correlation should not come as a surprise. Both the linear non-gravitational acceleration and the torques affecting spin angular momentum of a comet derive from the same physical process --- outgassing from the cometary surface. Thus, it is natural to expect that the two should be linked, which was noted by \citet{Whipple}. However, the focus of their work was on forced precession of the cometary spin driven by non-gravitational torques, which is not observable directly. For that reason, \citet{Whipple} had to assume a particular model of outgassing, in which $\boldsymbol{a}_{\rm ng}$ itself depends on the comet spin orientation, and to rely on $a_{\rm ng}$ measurements for comet 2P/Encke spanning about two centuries to deduce the evolution of spin orientation for this comet. Obviously, such extensive datasets are available for only for a handful of comets: \citet{Sekanina1984} used the same approach to study a century-long data set on comet P/Kopff.  

Our work is different from \citet{Whipple} in that we focus on the {\it period changes} of comets --- i.e. changes of the magnitude of the spin vector ${\bf S}$ rather than its orientation --- which have been  measured for a small number of objects directly. This makes our and \citet{Whipple}'s approaches complementary. A full treatment of the non-gravitational spin evolution of comets should consider both the magnitude and the orientation of ${\bf S}$ \citep{Neishtadt,Sidor}. 

However, even with our simple heuristic approach we are able to shed light on the non-gravitational torques acting on comets. In particular, using observational data (see \S \ref{sect:zeta}) we estimated in a model-independent fashion (i.e. not making any specific assumptions about the shape of the nucleus or geometry of outgassing) the effective dimensionless lever arm parameter $\zeta$, which relates the non-gravitational acceleration and the torque due to outgassing. This calculation assumes $\zeta$ to be constant for a given object and considers changes of only the spin rate, i.e. keeping spin axis fixed. These assumptions are likely somewhat simplistic. 

First, the geometry of Solar illumination changes as the comet moves along its orbit, likely modifying the distribution of active sites on the comet surface, which would then change both $a_{\rm ng}$ and the non-gravitational torques. As a consequence, parameter $\zeta$ defined by the equation (\ref{eq:F}) is expected to vary along the orbit in general. As our analysis focuses on the {\it integral} characteristics of spin evolution --- spin rate change per orbit and normalization $A$ of $a_{\rm ng}$ with a prescribed form of $g(r)$ (see equations (\ref{eq:a_ng}) and (\ref{eq:gr})) --- our results and statements regarding $\zeta$ effectively apply only to its orbit-averaged (in some complicated way) value. 

Second, forced precession of the rotational axis (neglected in our work) may significantly complicate the outgassing-driven spin evolution of comets, possibly making it somewhat stochastic (and better described as a random walk process rather than a simple linear trend in the most extreme cases). 

Despite these potential issues, based on the strong correlation established in \S \ref{sect:zeta} between the measured cometary spin period changes and theoretical predictions (Equations (\ref{eq:dOm1})-(\ref{eq:dOmmax})), we believe that our framework should still adequately describe the gross features of the cometary spin evolution.

It is interesting to compare the small values of $\zeta\lesssim 10^{-2}$ inferred in our work with other similar calculations. \citet{Whipple} and \citet{Sekanina1984} modeled comet nuclei as axisymmetric spheroids of small oblateness $\alpha\ll 1$, spinning around their (minor) polar axis, with axisymmetric (around spin axis) outgassing. In this particular model the dimensionless lever arm parameter (different from ours as it applies to forced precession) is $\zeta\approx \alpha\sin 2\psi$, where $\psi$ is the latitude of the sub-solar point relative to the spin axis ($0<\psi<\pi/2$). Assuming, very roughly, that $\psi$ (which varies as comet moves along its eccentric orbit) is uniformly distributed, one finds $\zeta\sim (2/\pi)\alpha$. \citet{Whipple} found $\alpha\approx 0.03$ for 2P/Encke, which would result in $\zeta\approx 0.02$, close to the upper limit implied by our estimate (\ref{eq:stats}), However, substantial rotational brightness modulation of the nucleus of 2P/Encke found by \citet{Luu} is at odds with such small a value of $\alpha$.

Using the same method, \citet{Sekanina1984} found high oblateness $\alpha\sim 0.15$ for P/Kopff, implying $\zeta\sim 0.1$ for this comet. This is substantially higher than our determinations of $\zeta$, implying that either this comet is an outlier of the distribution of $\zeta$, or that in some cases the model-dependent approach of \citet{Whipple} does not fully capture the details of cometary spin evolution.

\cite{Jewitt1997} evaluated outgassing torque and linear acceleration by modeling comet as a rectangular block with a single active site whose intensity is a harmonic function of the Solar illumination angle. Taking his results at face value we find $\zeta=2/\pi$, which is much larger than our findings (\ref{eq:stats}). For more realistic comet shapes and multiple active sites \cite{Jewitt1997} advocated smaller asymmetry of outgassing resulting in $\zeta$ of a few tens, which is still at least an order of magnitude larger than our measurements. Based on observations of spin dynamics of 9P/Tempel 1 \citet{Belton2011} suggested that asymmetry can be even lower (by about an order of magnitude), which would bring $\zeta$ into the range of values found in our work. 

A number of past studies focused on details of outgassing process, e.g. the dependence of mass loss and solar illumination on the cometary position along the orbit, detailed shape of the nucleus, its thermophysical properties, etc. However, majority of these investigations focused on either the linear non-gravitational acceleration \citep{Sosa,Sosa2011,Maquet,Hui} or the variation of the rotation period \citep{Keller,Mueller} separately, which is insufficient for determining $\zeta$. Our findings will hopefully provide motivation for future studies simultaneously accounting for both effects and their possible complicated interplay \citep{Whipple,Samar1995,KellerProc} that could provide a direct estimate of $\zeta$.

The small values of $\zeta$ found in this work are somewhat surprising, as they imply that the effective lever arm for application of non-gravitational force is very small compared to object's dimensions. In some cases this may be explained by geometric factors. For example, if the shape of the nucleus is close to spherical and outgassing is predominantly normal to the local surface (setup envisaged in \citep{Whipple} who found rather small effective lever arm for 2P/Encke), then the reactive force ${\bf f}$ should be very close to central at every point, giving rise to very weak torque. If, at the same time, the amplitude of the reactive force $|{\bf f}|$ strongly varies over the surface, then a substantial linear acceleration can be accumulated. This situation would naturally result in small $\zeta$, and could be relevant for some comets in our sample in Table \ref{table:zeta}. 

However, some comets, e.g. 67P/Churyumov-Gerasimenko, have highly non-spherical shapes, so that this argument would not work for them. In such cases smallness of $\zeta$ may argue in favor of a large number of active sites of outgassing randomly distributed over the cometary surface, resulting in a statistical cancellation (upon vector summation and averaging over the orbit) of the torques provided by them. The issue with this explanation may be that cancellation would also substantially reduce the amplitude of the net linear non-gravitational acceleration. However, if some systematic trend (e.g. proximity to sub-solar point) regulates the power of the outgassing vents, then $\boldsymbol{a}_{\rm ng}$ may be suppressed less than ${\bf T}$ upon vector summation over all outgassing sites, resulting in small $\zeta$. It may also be that, starting from an arbitrary rotational state, cometary spin ${\bf S}$ evolves in such a way over multiple approaches to the Sun as to minimize the net torque, something that is worth exploring further \citep{Neishtadt}.  

Relatively small spread of $\zeta$ that we found in this work --- about an order of magnitude (as $2\sigma_{\log\zeta}\approx 1$) for objects with $\Delta P$ spanning 4 decades --- also merits attention. To some degree, it could be a result of the small size of the sample (7 objects) used to derive  $\zeta$. On the other hand, there may be some other explanations, e.g. the aforementioned long-term spin evolution converging to some "attractor state" \citep{Neishtadt}.

This discussion makes it clear that measurements of $\zeta$ have a potential to help constrain the physics of cometary outgassing --- its geometry, intensity, etc. --- at least in a statistical sense. Better characterization of the distribution of $\zeta$ can have substantial impact on our understanding of cometary activity. This provides strong motivation for increasing the number of objects with measured $\Delta P$ and $a_{\rm ng}$. Right now the most basic properties of this distribution represented by equation (\ref{eq:stats}) are based on a sample of only 7 objects. Targeted observations of many other comets aimed at measuring their spin variations would allow us to characterize the distribution of $\zeta$ much better. According to Figure  \ref{fig:comets}, even period measurements with uncertainly of tens of minutes may reveal spin variability for dozens of comets with high $\Delta\Omega_1$ (the ones above the blue line in this figure). 

The objects with highest $\Delta\Omega_1$ listed in the Table \ref{table:fast} should have the best chance for revealing such changes, even though only one comet from Table \ref{table:zeta} (41P/Tuttle-Giacobini-Kres\'ak) has $\Delta\Omega_1>10^{-3}$ s$^{-1}$ necessary for being listed in Table \ref{table:fast}. Although we focused on studying the spin period variability of comets, non-gravitational forced precession  \citep{Whipple,Sekanina1984} should also be substantial for these objects, likely driving them into excited rotational state.


\subsection{Implications for Solar System comets}  
\label{sect:disc_SS}


Our calculations demonstrate that many Solar System comets (the ones with high $\Delta\Omega_1$) should exhibit rapid evolution of their rotation. Using equations (\ref{eq:spinev}) and (\ref{eq:a_ng}) one can estimate the characteristic timescale for the cometary spin evolution $\tau_\Omega=|\Omega/\dot\Omega|$ as 
\ba
\tau_\Omega &=& \frac{4\pi}{5}\zeta^{-1}\frac{R}{PAg(r)}
\label{eq:tau_Om}\\
&\approx & 670~\mbox{d}~\left[g(r)\right]^{-1}\frac{R}{1~\mbox{km}}
\nonumber\\
&\times & \left(\frac{\zeta}{0.006}~\frac{A}{10^{-8}\mbox{au d}^{-2}}~\frac{P}{10~\mbox{hr}}\right)^{-1}.
\ea   
This timescale is comparable to the time comets spend in the inner Solar System (about a year), implying that objects with $A\gtrsim 10^{-8}$ au d$^{-2}$ (quite numerous in Figure \ref{fig:comets}) and perihelia within $1$ au should experience substantial changes of their period in a single orbit. Moreover, comets with $A\gtrsim 10^{-7}$ au d$^{-2}$ could be spun up to the limit of rotational breakup in a single orbit even starting from a non-spinning state. Of course, this estimate assumes that the lever arm parameter $\zeta$ of such comets obey the same distribution as for objects studied in \S \ref{sect:zeta}. 

In reality, situation is not as dramatic since many comets with high $A$ have perihelia outside 1 au, meaning $g(r)\lesssim 1$ and longer $\tau_\Omega$. Indeed, only three objects discussed in \S \ref{sect:comets} are close to the "breakup" (red) line in Figure \ref{fig:comets}  as predicted by Equations (\ref{eq:dOm1})-(\ref{eq:dOmmax}). 

Nevertheless, our results do suggest that rotational fission triggered by the non-gravitational torques is an important destruction mechanism for many comets, as advocated by many authors in the past \citep{Samar1995,Jewitt2016,Jewitt2017}. Our adoption of $P_{\rm crit}=1$ hr as a period leading to rotational breakup is rather conservative. Internally weak comets are likely to fission at longer spin periods. 

On the other hand, our estimates of $\Delta\Omega$ based on Equations (\ref{eq:dOm1})-(\ref{eq:dOmmax}) assume that non-gravitational torques always spin up comets in the same direction, keeping the rotational axis fixed. As mentioned before, this assumption is likely only a rough approximation; resulting departures from a simple linear trend in $\Omega$ would slow down the spin-up of comets towards rotational fission.

Possibility of rotational fission provides an interesting indirect constraint on the values of the lever arm parameter $\zeta$. In Figure \ref{fig:comets} there is only one object  above the (upper red dotted) line corresponding to spin-up to $P_{\rm crit}=1$ hr with $\zeta=0.0017$ (i.e. $\zeta=10^{\langle\log\zeta\rangle-\sigma_{\log\zeta}}$), and this object is C/2015 D1 (SOHO), which did not survive its perihelion passage, see \S \ref{sect:comets}. If many comets had substantially smaller values of $\zeta$, then there would be no reason for them to not populate the region $\Delta\Omega_1>1$ s$^{-1}$: given low enough $\zeta$, even such large values of $\Delta\Omega_1$ would still result in low enough $\Delta\Omega$ to avoid rotational fission, see equation (\ref{eq:dOm1}). Our interpretation of the lack of comets with $\Delta\Omega_1>1$ s$^{-1}$, most susceptible to rapid rotational evolution, is that there is a {\it lower limit} on the values of their lever arm parameter, $\zeta\gtrsim 10^{-3}$, so that all objects with $\Delta\Omega_1$ get spin up to breakup in a single passage through the inner Solar System. In other words, the lack of comets with high values of $\Delta\Omega_1$ is due to a "survival bias" caused by the important role of rotational fission promoted by outgassing. This bias may also explain small values of $\zeta$ found in our work, as objects with large $\zeta$ rapidly spin up and get preferntially destroyed by rotational fission.


\section{Summary}  
\label{sect:summary}


We have investigated the relation between the non-gravitational linear acceleration of comets and their spin evolution driven by outgassing. In our work we focused on spin period changes caused by the non-gravitational torques, as opposed to other studies concentrating on forced precession of the spin axis \citep{Whipple,Sekanina1984}. Our main conclusions are listed below. 

\begin{enumerate}

\item Based on heuristic arguments we proposed a simple linear relation between the variation of cometary spin period (and rate) and the net non-gravitational acceleration. This relation depends on a single parameter --- (dimensionless) effective lever arm $\zeta$, linking the non-gravitational torque and acceleration.

\item Using a sample of 7 comets with measured $a_{\rm ng}$ and spin period changes we verified the validity of this relation. We also measured values of $\zeta$ for these objects and found $\log\zeta=-2.21\pm 0.54$ for the whole sample.

\item Using our framework we computed expected changes of spin period (per orbit) for a much larger sample of comets (209 objects) with measured non-gravitational accelerations. Assuming the inferred distribution of $\zeta$ to hold for these comets, we showed that several dozens of them should exhibit spin period changes per orbit of order an hour. A handful of comets may be in danger of rotational fission because of their rapid spin-up due to outgassing torques. This process must be an important channel of cometary destruction.
    
\item We advocate the use of $\Delta\Omega_1$ --- change of the cometary spin rate in a single orbit evaluated for lever arm parameter $\zeta=1$, defined by the equation (\ref{eq:dOmmax}) --- as a metric for assessing the potential of a particular object to exhibit large changes of spin period. This parameter can be used for guiding target selection for measurements of cometary spin variability, which would help us better constrain the true distribution of $\zeta$.
    
\end{enumerate}

Our results will assist future observations of spin period variability of comets, helping us better understand physics of cometary outgassing. They could also be relevant for spin dynamics of other minor objects, such as asteroids, both in the Solar System \citep{Jewitt2017} and beyond (R. Rafikov, 2018, in preparation).

\acknowledgements

Financial support for this study has been provided by NSF via grant AST-1409524 and NASA via grant 15-XRP15-2-0139.



\bibliographystyle{apj}
\bibliography{references}

\begin{table*}
\begin{threeparttable}
\caption{Solar System objects with measured non-gravitational acceleration and spin period changes}
\begin{tabular}{lccccccc}
    \hline \hline\\
    Name & $a$ & $e$ & $R$ & $A_r$ $(A_1)$ & $A_\varphi$ $(A_2)$ & $A_n$ $(A_3)$  & $\Delta \Omega_1$ 
    \\ 
     & $\big[$au$\big]$ &  & $\big[$km$\big]$ & $\big[$au d$^{-2}\big]$ & $\big[$au d$^{-2}\big]$ & $\big[$au d$^{-2}\big]$  & $\big[$s$^{-1}\big]$  \\
    \\
    \hline
    \\
 C/2015 D1 (SOHO)\footnotemark[1]  &	4.94  &	0.99427  &	- 	& $1.25\times 10^{-6}$ &	0  &	0  &	2.36\\ 
     P/1999 J6 (SOHO)  &	3.1  &	0.9842  &	-	& $3.9\times 10^{-7}$	& $-1.89\times 10^{-9}$ &	0  &	0.518\\ 
  205P/Giacobini-B  &	3.60  &	0.575  &	-	& $7.8\times 10^{-6}$	& $-2.4\times 10^{-7}$ &	0  &	0.486\\ 
     C/2017 E4 (Lovejoy)  &	477.7  &	0.998967  &	-	& $1.74\times 10^{-7}$	& $-7.46\times 10^{-8}$ &	0  &	0.0424\\ 
   86P/Wild 3  & 	3.62  &	0.369  &	0.86	& $3.1\times 10^{-7}$	& $5.7\times 10^{-8}$ &	0  &	0.0405\\ 
     C/2012 V1 (PANSTARRS)  &	3785.8  &	0.99945  &	-	& $2.2\times 10^{-6}$	& $1.1\times 10^{-7}$	& $2.5\times 10^{-7}$ &	0.027\\ 
     C/1996 B2 (Hyakutake)  &	2272.1  &	0.99989867  &	4.2	& $2.3\times 10^{-8}$	& $3.6\times 10^{-10}$ &	0  &	0.0231\\ 
     C/1998 P1 (Williams)  &	1698.4  &	0.999325 &	-	& $2.69\times 10^{-7}$	& $1.46\times 10^{-8}$ &	0  &	0.0227\\ 
   41P/Tuttle-Giacobini-Kresak  &	3.085 &	0.661  &	1.4	& $1.74\times 10^{-8}$	& $4.29\times 10^{-9}$	& $1.47\times 10^{-9}$ &	0.0159\\ 
   73P/Schwassmann-Wachmann 3-BT  &	3.09  &	0.686  &	-	& $-1.05\times 10^{-7}$	& $3.11\times 10^{-8}$ &	0  &	0.0148\\ 
  147P/Kushida-Muramatsu  &	3.807  &	0.276  &	0.42	& $2.67\times 10^{-7}$	& $-6.9\times 10^{-8}$	& $1.63\times 10^{-7}$ &	0.0140\\ 
     C/2002 X5 (Kudo-Fujikawa)  &	1208.7  &	0.99984  &	-	& $2.52\times 10^{-8}$	& $5.83\times 10^{-9}$ &	0  &	0.0127\\ 
  104P/Kowal 2  &	3.263  &	0.639 &	2	& $2.2\times 10^{-8}$	& $-1.06\times 10^{-10}$	& $4.3\times 10^{-9}$ &	0.0117\\ 
     C/1987 U3 (McNaught)	  & 576.5 &	0.99854  &	-	& $8.23\times 10^{-8}$	& $3.8\times 10^{-9}$ &	0  &	0.0107\\ 
   71P/Clark  &	3.137  &	0.494  &	1.36	& $2.12\times 10^{-8}$	& $8.65\times 10^{-9}$ &	0  &	0.0102\\ 
   27P/Crommelin  &	9.231  &	0.91898  &	-	& $-3.66\times 10^{-8}$	& $-1.71\times 10^{-9}$	& $2.08\times 10^{-9}$ &	0.0057\\ 
   87P/Bus	  & 3.488  &	0.376  &	0.56	& $2.0\times 10^{-8}$	& $-2.33\times 10^{-9}$	& $1.15\times 10^{-9}$ &	0.0057\\ 
  141P/Machholz 2-D	 & 3.009  &	0.751  &	-	& $2.27\times 10^{-8}$	& $5.86\times 10^{-9}$ &	0  &	0.00415\\ 
     C/2001 A2-A (LINEAR)	  & 2530.5 &	0.999692 &	-	& $-2.11\times 10^{-8}$	& $1.92\times 10^{-8}$ &	0  &	0.00404\\
   51P/Harrington  &	3.714  &	0.542  &	4.8	& $3.87\times 10^{-8}$	& $4.0\times 10^{-9}$	& $-7.13\times 10^{-9}$ &	0.00377\\ 
   75D/Kohoutek  &	3.543  &	0.496  &	4.6	& $3.98\times 10^{-8}$	& $8.7\times 10^{-9}$ &	0  &	0.0035\\ 
     C/2014 E2 (Jacques)  &	688.3  &	0.99904  &	-	& $2.07\times 10^{-8}$	& $-2.82\times 10^{-9}$ &	0  &	0.00351\\ 
     C/2014 Q1 (PANSTARRS)  &	1129.4  &	0.99972  &	-	& $8.21\times 10^{-9}$	& $2.75\times 10^{-9}$	& $-5.56\times 10^{-9}$ &	0.00339\\ 
   51P/Harrington-A  &	3.581  &	0.562  &	4.8	& $2.63\times 10^{-8}$	& $3.1\times 10^{-9}$ &	0  &	0.00322\\ 
    5D/Brorsen  &	3.101 & 	0.81  &	-	& $1.27\times 10^{-8}$	& $1.34\times 10^{-9}$ &	0  &	0.00278\\ 
   73P/Schwassmann-Wachmann 3-E  &	3.062  &	0.694 &	-	& $1.72\times 10^{-8}$	& $5.81\times 10^{-9}$ &	0  &	0.00256\\ 
     P/2007 T2 (Kowalski)  &	3.0927  &	0.775  &	-	& $1.32\times 10^{-8}$	& $2.31\times 10^{-9}$ &	0  &	0.00252\\ 
     C/1999 J3 (LINEAR)  &	1596.1  & 	0.999388  &	-	& $2.24\times 10^{-8}$	& $-5.42\times 10^{-9}$ &	0  &	0.00246\\ 
     C/2002 V1 (NEAT)  &	1010.7  &	0.999902  &	-	& $-2.96\times 10^{-9}$	& $3.79\times 10^{-10}$ &	0  &	0.00238\\ 
   21P/Giacobini-Zinner  &	3.499  &	0.71  &	2	& $3.74\times 10^{-9}$	& $-1.04\times 10^{-10}$ &	0  &	0.00232\\ 
     C/1985 R1 (Hartley-Good)  &	5982.4  &	0.999884  &	-	& $1.33\times 10^{-8}$	& $-2.16\times 10^{-9}$ &	0  &	0.00217\\ 
  316P/LONEOS-Christensen  &	4.328  &	0.166  &	-	& $7.34\times 10^{-5}$ &	0  &	0  &	0.00212\\ 
   73P/Schwassmann-Wachmann 3-B  &	3.062  &	0.693  &	-	& $1.49\times 10^{-8}$	& $1.96\times 10^{-9}$ &	0  &	0.00212\\ 
     C/1993 Y1 (McNaught-Russell)  &	134.76  &	0.99356 &	 -	& $1.65\times 10^{-8}$	& $1.22\times 10^{-9}$ &	0  &	0.00207\\ 
     C/1999 T1 (McNaught-Hartley)  &	8149.7  &	0.999856  &	-	& $2.44\times 10^{-8}$	& $-9.59\times 10^{-10}$ &	0  &	0.00198\\ 
  101P/Chernykh  &	5.785  &	0.594  &	5.6	& $-5.16\times 10^{-8}$	& $1.55\times 10^{-7}$ &	0  &	0.00197\\ 
  168P/Hergenrother  &	3.624  &	0.61  &	-	& $2.66\times 10^{-8}$	& $2.37\times 10^{-9}$	& $-1.25\times 10^{-9}$ &	0.00196\\ 
   76P/West-Kohoutek-Ikemura  &	3.471  &	0.539  & 	0.66	& $-2.18\times 10^{-9}$	& $-1.85\times 10^{-10}$ &	0  &	0.00186\\ 
  252P/LINEAR  &	3.047  &	0.673  &	-	& $1.05\times 10^{-8}$	& $-7.06\times 10^{-9}$ &	0  &	0.00167\\ 
   88P/Howell  &	3.11  &	0.562  &	4.4	& $8.41\times 10^{-9}$	& $-1.6\times 10^{-9}$	& $-5.2\times 10^{-10}$ &	0.00164\\ 
     C/2011 F1 (LINEAR)  &	2776.2  &	0.999345  &	-	& $5.67\times 10^{-8}$	& $-2.5\times 10^{-8}$	& $-1.47\times 10^{-9}$ &	0.00157\\ 
   59P/Kearns-Kwee  &	4.485  &	0.475  &	1.58	& $3.24\times 10^{-8}$	& $-1.88\times 10^{-9}$	& $2.1\times 10^{-9}$ &	0.00154\\ 
  154P/Brewington  &	4.883  &	0.671  &	-	& $2.9\times 10^{-8}$	& $-2.02\times 10^{-9}$	& $-9.65\times 10^{-9}$ &	0.0015\\ 
     C/2012 X1 (LINEAR)  &	153.3 &	0.98957  & 	-	& $3.61\times 10^{-8}$	& $2.19\times 10^{-9}$	& $5.33\times 10^{-9}$ &	0.00148\\ 
  240P/NEAT  & 	3.866  &	0.45  &	-	& $7.89\times 10^{-8}$	& $3.47\times 10^{-8}$	& $9.91\times 10^{-10}$ &	0.00145\\ 
  205P/Giacobini-A  &	3.539  &	0.569  &	-	& $1.38\times 10^{-8}$	& $-1.72\times 10^{-8}$ &	0  &	0.00138\\ 
  141P/Machholz 2  &	3.019  &	0.749  &	-	& $7.64\times 10^{-9}$	& $2.21\times 10^{-10}$ &	0  &	0.00134\\ 
    2P/Encke  &	2.215  &	0.848  &	4.8	& $-1.34\times 10^{-11}$	& $-2.28\times 10^{-12}$	& $1.8\times 10^{-9}$ &	0.00133\\ 
   42P/Neujmin 3  &	4.876  &	0.584  &	2.2	& $1.27\times 10^{-8}$	& $-1.05\times 10^{-9}$ &	0  &	0.00113\\ 
  255P/Levy  &	3.038  &	0.668  &	-	& $7.33\times 10^{-9}$	& $-4.15\times 10^{-9}$ &	0  &	0.0011\\ 
 \\
    \hline   
\end{tabular}
\begin{tablenotes}
\item {\bf Notes}: For every comet we list semi-major axis and eccentricity of its heliocentric orbit, effective physical radius (if available), 3 components of the non-gravitational acceleration at 1 au 
$(A_r,A_\varphi,A_n)=(A_1,A_2,A_3)$ assuming \citet{Marsden} model (\ref{eq:gr}), and $\Delta\Omega_1$ --- theoretical prediction for the change of spin rate over an orbit for a lever arm parameter $\zeta=1$ defined by equation (\ref{eq:dOmmax}). Data on $(A_r,A_\varphi,A_n)$ come from Small Body Database (https://ssd.jpl.nasa.gov/?comets).
\end{tablenotes}
\label{table:fast}
\end{threeparttable}
\end{table*}

\end{document}